\documentclass[a4paper,10pt,twoside]{cpc-hepnp}
\usepackage[numbers,sort&compress]{natbib}
\usepackage{stmaryrd}
\usepackage{bbding}
\usepackage{amssymb}
\usepackage{multicol}
\usepackage{multirow}
\usepackage{graphicx}
\usepackage{subfigure}
\usepackage{booktabs}
\usepackage{amssymb,bm,mathrsfs,bbm,amscd}
\usepackage[tbtags]{amsmath}
\usepackage{lastpage}
\usepackage{CJK}



\begin{document}
\begin{CJK*}{GBK}{song}

\fancyhead[co]{\footnotesize Xiu Qing-Lei et al: Study of the Tracking Method and Expected Performance of the Silicon Pixel Inner Tracker Applied in BESIII}


\title{Study of the Tracking Method and Expected Performance of the Silicon Pixel Inner Tracker Applied in BESIII
\thanks{Supported by Joint Funds of National Natural Science Foundation of China (U1232202) and National Natural Science Foundation of China (11205184, 11205182)}}

\author{%
\quad Qinglei Xiu$^{1,2,3;1}$\email{xiuql@ihep.ac.cn} %
\quad Mingyi Dong$^{1,2}$%
\quad Weidong Li$^{2}$ \\
\quad Huaimin Liu$^{2}$ %
\quad Qiumei Ma$^{2}$ %
\quad Qun Ouyang$^{1,2}$ %
\quad Zhonghua Qin$^{1,2}$ \\
\quad Liangliang Wang$^{2}$ %
\quad Linghui Wu$^{2}$ %
\quad Ye Yuan$^{2}$ %
\quad Yao Zhang$^{2}$
}
\maketitle

\address{%
$^1$ State Key Laboratory of Particle Detection and Electronics, Beijing 100049, China\\
$^2$ Institute of High Energy Physics, CAS, Beijing 100049, China\\
$^3$ University of Chinese Academy of Sciences, Beijing 100049, China\\
}

\begin{abstract}
The inner drift chamber of the BESIII is encountering serious aging problem after five year's running. For the first layer, the decrease in gas gain is about 26\% from 2009 to 2013. The upgrade of the inner tracking detector has become an urgent problem for the BESIII experiment. An inner tracker using CMOS pixel sensors is an important candidate because of its great advantages on spatial resolution and radiation hardness. In order to carry out a Monte Carlo study on the expected performance, a Geant4-based full simulation for the silicon pixel detector has been implemented. The tracking method combining the silicon pixel inner tracker and outer drift chamber has been studied and a preliminary reconstruction software was developed. The Monte Carlo study shows that the performances including momentum resolution, vertex resolution and the tracking efficiency are significantly improved due to the good spatial resolution and moderate material budget of the silicon pixel detector.
\end{abstract}



\begin{multicols}{2}

\section{Introduction}
The main drift chamber(MDC) is the center tracking detector of the Beijing spectrometer III(BESIII)\cite{bes3} which is operating at the Beijing electron positron collider II(BEPCII)\cite{bepc2}. In order to meet the requirements of BESIII experiment, MDC is designed to be a small-cell, low-mass drift chamber using a helium-based gas mixture. The drift chamber contains an inner chamber which consists of 8 stereo layers with the drift cell in the size of about $12 \times 12 mm^2$, and an out chamber which consists of 35 layers with the drift cell in the size of about $16.2 \times 16.2 mm^2$. The inner chamber was designed to be replaceable in consideration of radiation damage.

Because of high beam related background, the inner chamber is suffering from aging problem after it has been running five years, and the gain is dropping year by year. For the first layer, the gain decreases by 26\% from 2009 to 2013. Thus, it is necessary to make preparations for the upgrade of MDC inner chamber. The silicon pixel tracker(SPT) using CMOS pixel sensors(CPS)\cite{cps_Turchetta} which is first developed by IPHC for charged particle tracking is a good candidate because of the low material budget($\approx 50 \mu m $), good spatial resolution($< 10 \mu m$) and high radiation tolerance. In order to do a Monte Carlo study on the expected performance of applying SPT in BESIII,  a Geant4-based full simulation and a track reconstruction software have been developed in the BESIII Offline Software System (BOSS)\cite{boss}. In this paper we introduce the study of the tracking method combining SPT and MDC outer chamber. Expected performances including tracking efficiency, momentum resolution and vertex resolution from Monte Carlo study are also presented.

\begin{figure*}[htb]
\centering

\subfigure[$R-\Phi$ view]{\includegraphics[height=5cm]{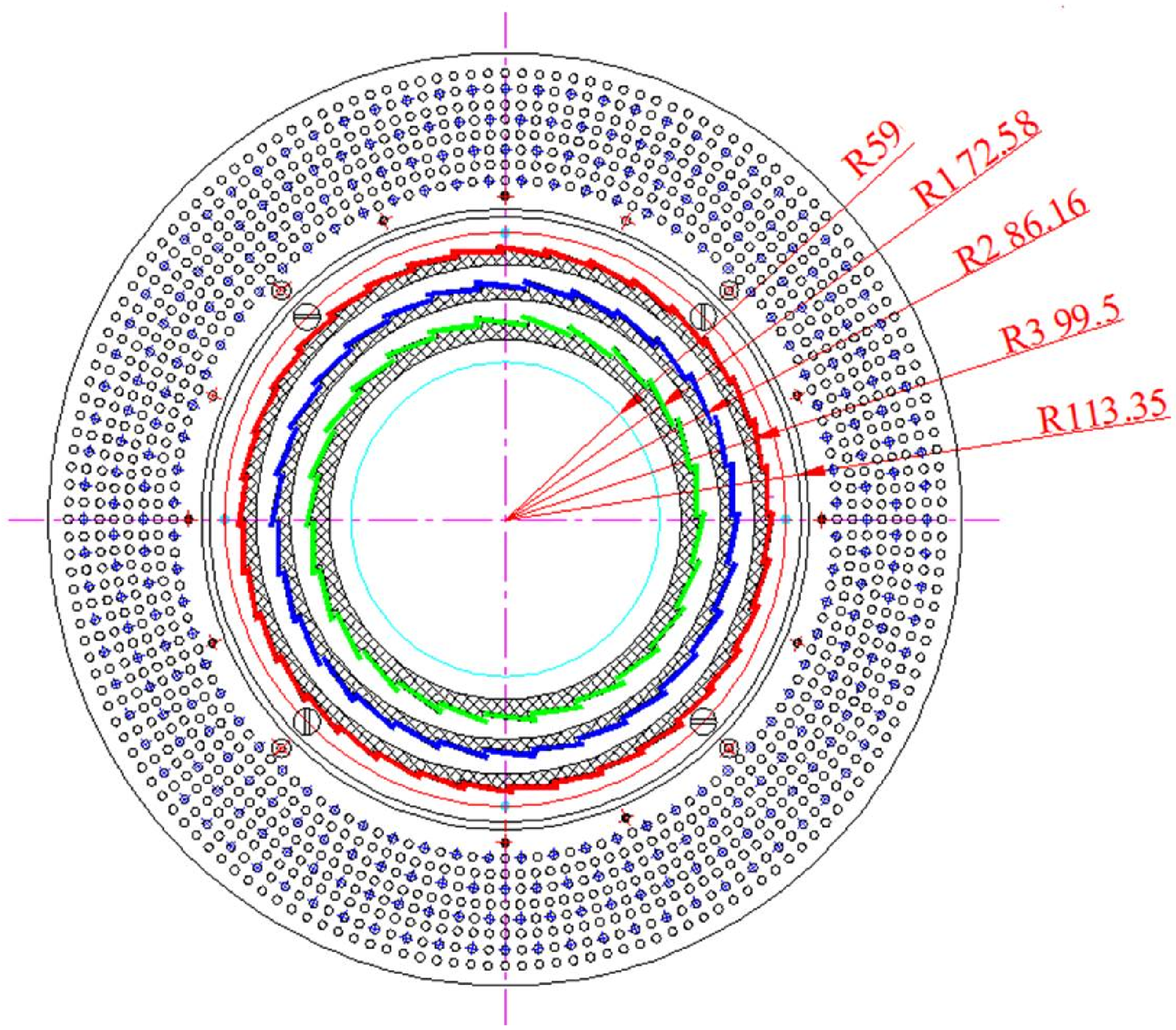}\label{subfig:sptDesignRPhi}}
\subfigure[Z view]{\includegraphics[height=5cm]{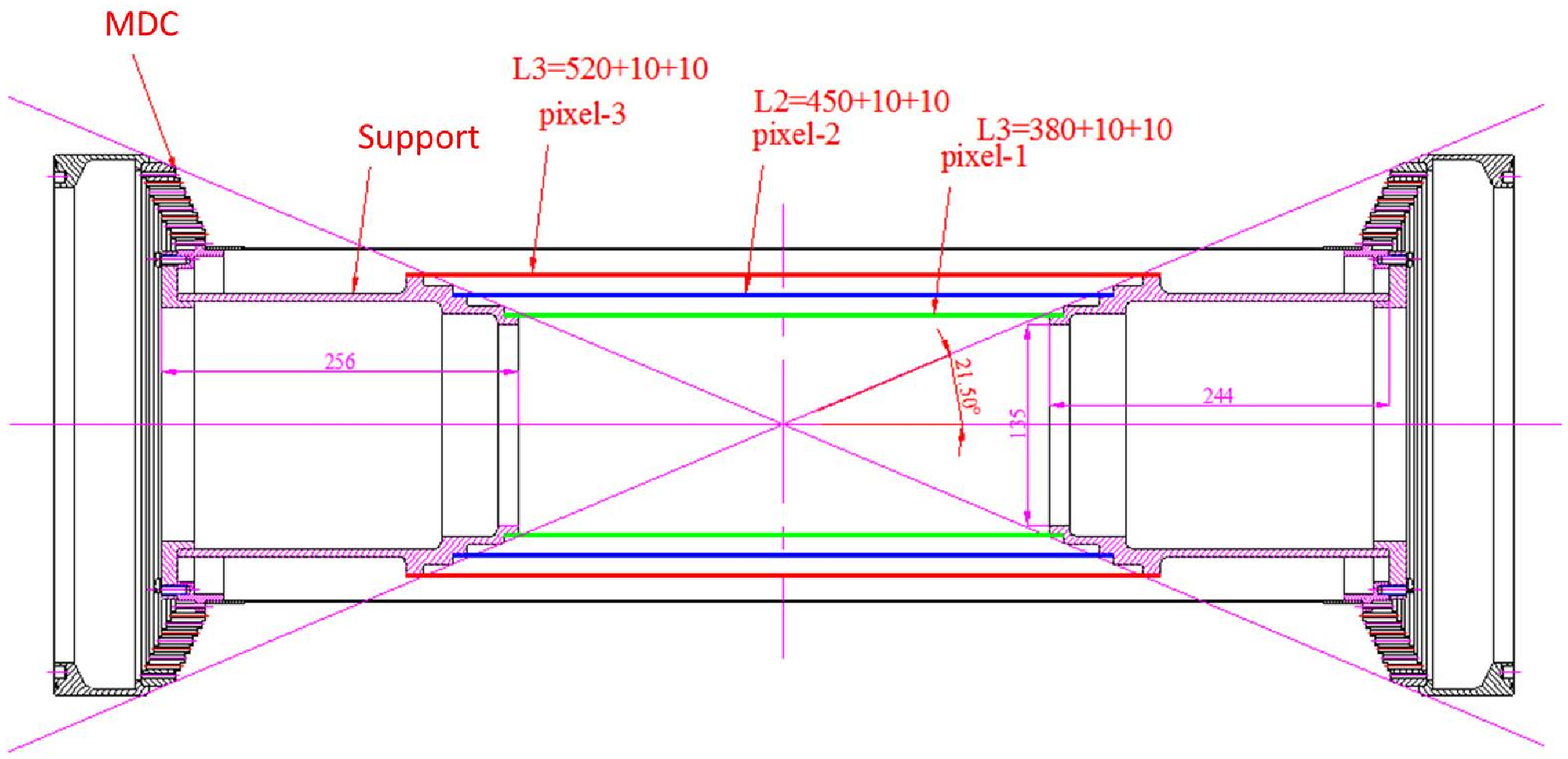}\label{subfig:sptDesignZ}}
\caption{Preliminary design of SPT}

\label{fig:sptDisign}
\end{figure*}

\begin{table*}[htb]
\centering
\caption{Material Budget of one layer in SPT}
\begin{tabular}{|c|c|c|c|}
\hline
\multicolumn{2}{|c|}{Structure} & Thickness($\mu m$) & Radiation Length($ X / X_0 $)\\
\hline
\multicolumn{2}{|c|}{Chip} & 50 & 0.05\% \\
\hline
\multicolumn{2}{|c|}{Epoxy} & 50 & 0.014\% \\
\hline
\multirow{2}{*}{Flex Cable} & Kapton & 100 & 0.035\%  \\
\cline{2-4}
& Aluminum & 50 & 0.056\% \\
\hline
\multicolumn{2}{|c|}{Epoxy} & 50 & 0.014\% \\
\hline
\multicolumn{2}{|c|}{Carbon Fiber} & 500 & 0.21\% \\
\hline
\multicolumn{2}{|c|}{Total} & 800 & 0.38\% \\
\hline
\end{tabular}

\label{tab:sptMateiralBudget}
\end{table*}

\section{The preliminary design of SPT}

The inner drift chamber of MDC is preliminarily designed to be replaced by 3 layers of silicon pixel detectors at the radius of 72.58mm, 86.16mm and 99.5mm, with the length of 380mm, 450mm and 520mm arranged from the inner layer to the outer layer(Figure \ref{fig:sptDisign}).

Every layer includes $26 \sim 36$ ladders which are the basic modules of SPT and fixed at MDC endplates. There is an overlap region of about 10\% between two ladders in $R-\Phi$ direction to eliminate the dead area. Each ladder, with different components connected by glue, consists of mechanical supports made of carbon fiber, readout cable made from kapton coated with aluminum  and $20 \sim 30$ CPS chips on the top.

The material budget of one layer shown in Table \ref{tab:sptMateiralBudget} is totally 0.38\% radiation length, which is mainly caused by 500$\mu m$ thick carbon fiber support.

The description of the geometry and materials of SPT has been implemented in the Geant4-based full simulation package in BOSS according to the preliminary design, and a simple digitization model is established to simulate the charge collection in the CPS chips.

\section{Track reconstruction of SPT}
For the current MDC tracking, the charged tracks are reconstructed by the pattern matching method\cite{pat} combined with the conformal transformation method\cite{tsf}. However, these reconstruction methods can't work in SPT when the new silicon inner tracker is applied in BESIII. Thus, new reconstruction methods have to be developed for the SPT. In this study, a new method called combinatorial Kalman filter(CKF)\cite{ckf}\cite{kf_fruhwirth}\cite{adaptiveFilter} is used to develop the track reconstruction software for SPT. The procedure to reconstruct tracks in one event could be divided into three steps:
 \begin{itemize}
  \item Firstly, all possible track seeds will be found in one sub-detector
  \item Then, a set of track candidates will be built for each seed by extrapolate the seed into other sub-detectors layer by layer.
  \item Finally, the best candidate for each seed will be selected and seeds from the same track will also be merged.
\end{itemize}

Actually, there are two complementary sequences to reconstruct the track for the CKF method. One is the outside-in sequence which find seeds in MDC outer chamber and extrapolate the seeds into SPT. It's implied that the SPT is just used to update the tracks found in MDC outer chamber but not used to find tracks. Thus, the track efficiency of this method will be lower than the full MDC. The other is the inside-out sequence which find seeds in SPT and extrapolate the seeds into MDC outer chamber. In order to increase the track efficiency, both of these two sequences are implemented in our software and the track candidates found by these two sequences separately will be merged by comparing the hits of the track candidates. The software is developed on the foundation of BESIII track fitting algorithm\cite{bes_fit}, in which the basic Kalman filter for MDC has been implemented.

\subsection{Outside-in track reconstruction}

The first step in the outside-in track reconstruction is to find track seeds in the MDC outer chamber by the pattern matching method combined with the conformal transformation method.

Then these tracks will be extrapolated into SPT to match the hits by iterating the following steps:
\begin{itemize}
  \item Extrapolate all track candidates to next layer in the prediction step of kalman filter
  \item Look for compatible hits around the predicted point for each candidate
  \item For every candidate, generate a branch for each compatible hit and update the branch with the hit.
  \item Drop bad candidates according to the number of missing hits, $\chi^2$ and so on.
\end{itemize}

After all of the hits have been processed, the best candidate will be selected for each seed according to the number of hits and $\chi^2$.

Figure \ref{fig:spt_inn_extr} shows the iterative process of the outside-in sequence in the SPT. Track candidates in SPT will be extended toward the interaction point layer by layer. In each layer, the multiple-scattering and energy loss will be considered for each candidate. Because there is a little bit of overlap region between two ladders in one layer, the created branch candidates will be tested whether intersect with other ladders in this layer. If another ladder intersects with the branch candidate, the iterative process will be invoked again.

\begin{center}
\includegraphics[width=8cm]{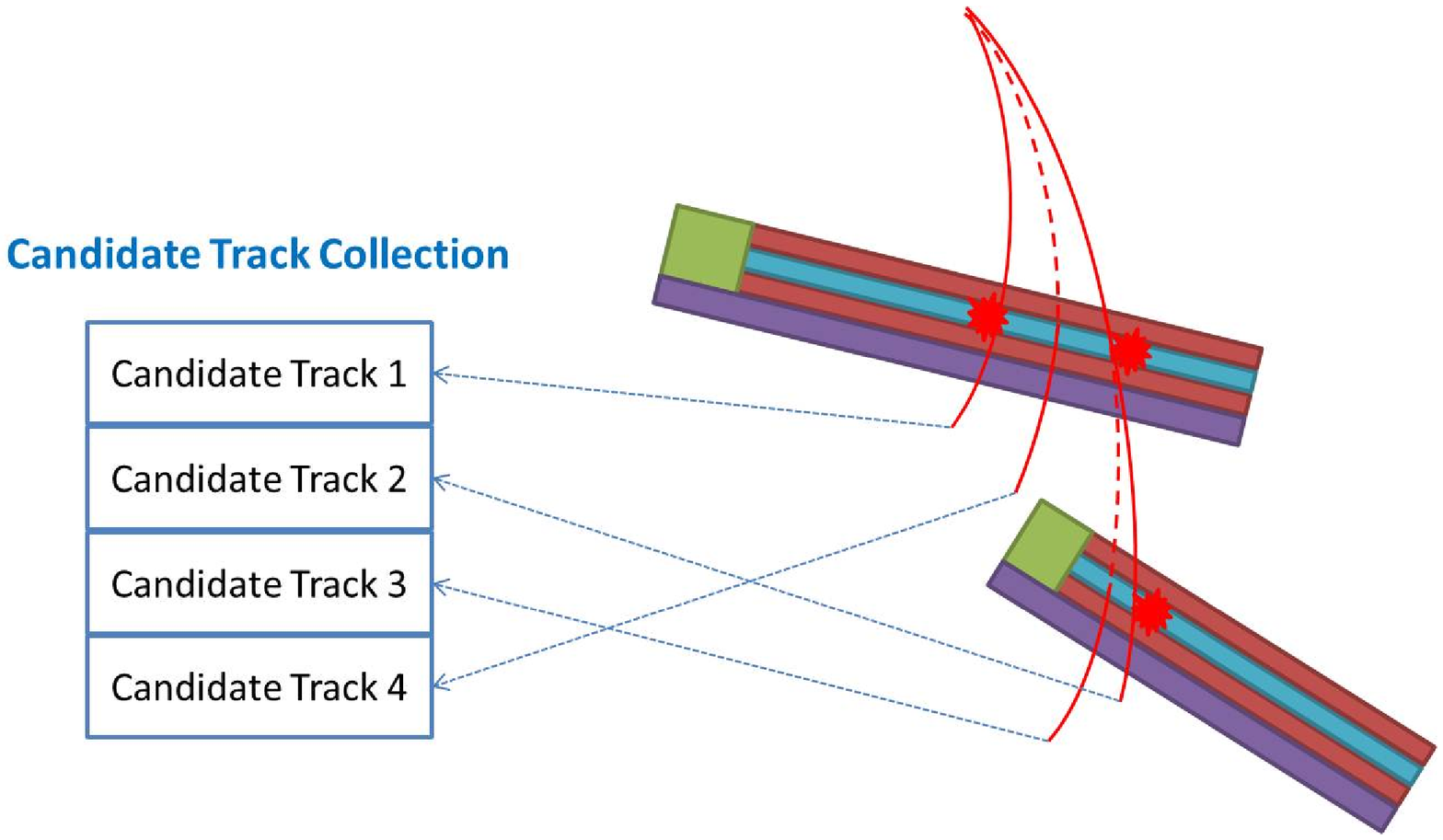}
\figcaption{\label{fig:spt_inn_extr}A tree of candidates are build up in a layer from one candidate. The dashed lines are the prediction of the original candidate and the solid lines are the updated branch candidates}
\end{center}

\subsection{Inside-out track reconstruction}
Since the track efficiency of the outside-in sequence would be lower than the full MDC, a complementary inside-out sequence is necessary. The inside-out tracking algorithm starts from the track seeds found in SPT. The iterative process of the inside-out algorithm is very similar with that of the outside-in algorithm except that the propagate direction is from SPT to MDC.

In order to obtain higher tracking efficiency, the track seed coming from the colliding point consists of two compatible hits(hit pair) at different layers in SPT with a loose constraint of the beam spot. In the hit pairs finding, all hits at outer layer with lager radius will be iterated. For each outer hit, the beam spot constraint and the minimum transverse momentum cut are applied to estimate a window in $\Phi$ direction in the inner layer with a smaller radius. Hit pairs will be created for each hit in the $\Phi$ window of the inner layer. Because 3 layers are included in SPT, there will be 3 kinds of combination of layers in the hit pair finding. Figure \ref{fig:hit_pair} shows the hit pair finding in two layers.

\begin{center}
\includegraphics[width=5cm]{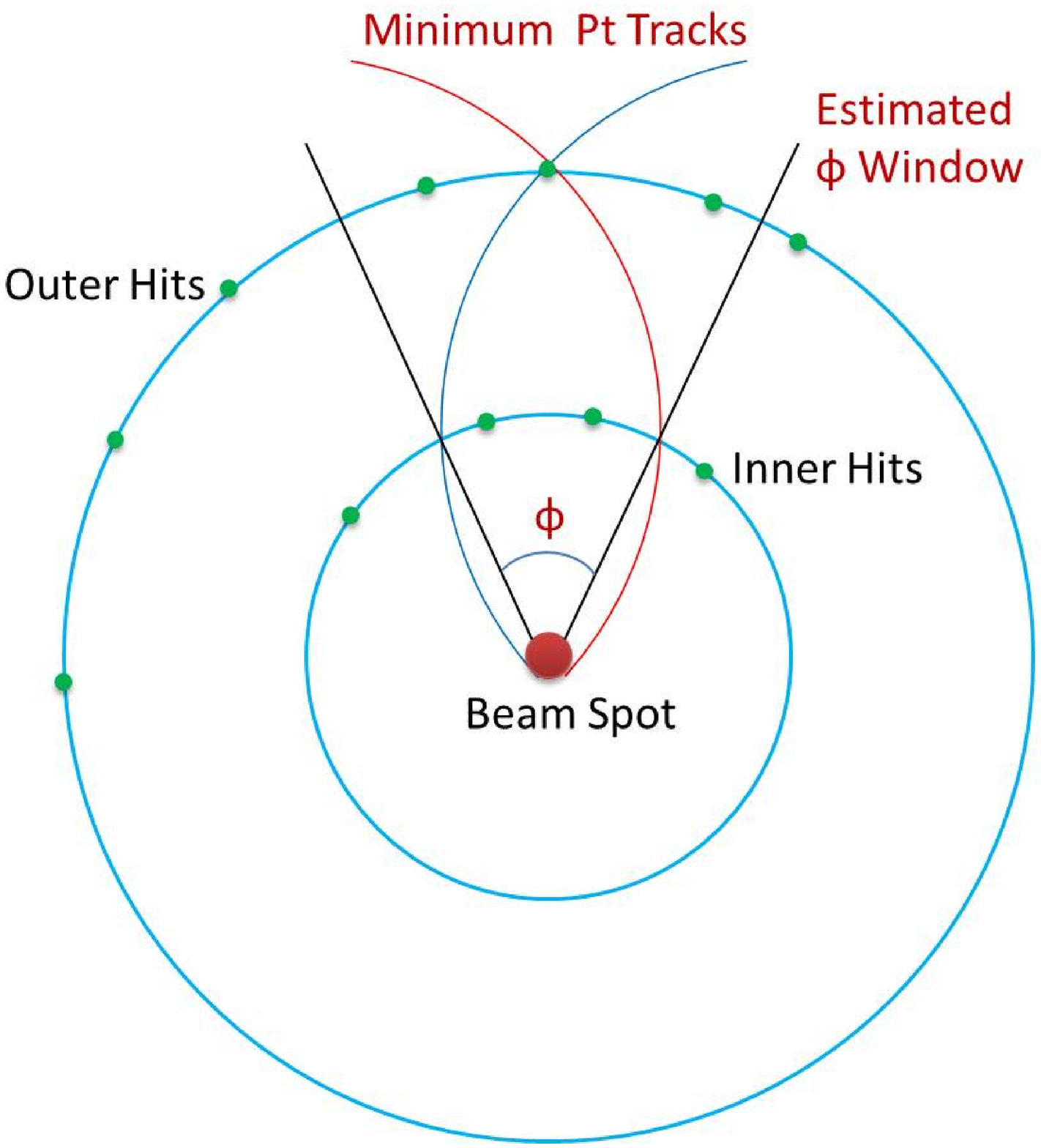}
\figcaption{\label{fig:hit_pair}Illustration of the hit pair finding}
\end{center}

\section{Expected performance}
Monte Carlo single muon tracks are used to study the expected tracking performances, including momentum resolution, vertex resolution and tracking efficiency. We also generate $\Psi(2S) \to \pi\pi J/\psi$ samples to check the invariance mass distribution. In this section, the results of current MDC is labeled as MDC and  the results of applying SPT inner tracker is labeled as SPT for convenience.

\subsection{The momentum resolution}
The momentum resolution is defined as $\sigma_{p} / p$, which is mainly influenced by the spatial resolution and the multiple scattering. Figure \ref{fig:resP}, which gives the momentum resolution as a function of the track momentum, shows the improvement of the momentum resolution in high momentum region after applying SPT inner tracker. For 1GeV/c tracks, the momentum resolution is improved from 0.53\% to 0.46\%. However, in low momentum region, the improvement is very small because more material budget in SPT results to more contribution from multiple scattering effect.

\begin{center}
\includegraphics[width=8cm]{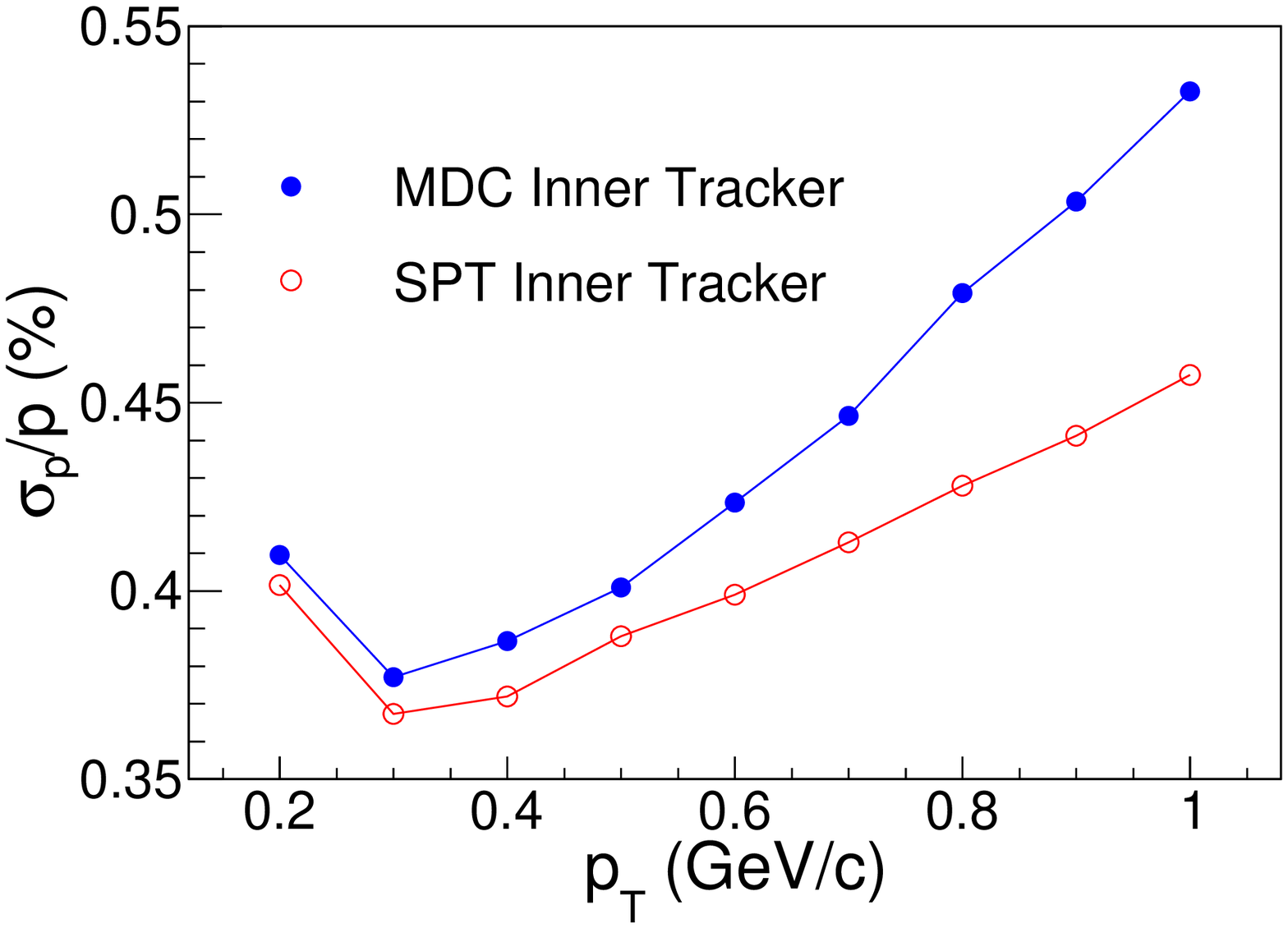}
\figcaption{\label{fig:resP}The momentum resolution as a function of the momentum}
\end{center}

\subsection{The vertex resolution}
The information of the vertex can be obtained from the track parameter $d \rho$ and $dz$. We define the residual as the difference between the reconstructed value and MC truth value. The vertex resolution can be represented by $\sigma(rho)$ and $\sigma(z)$, which are obtained from fitting a Gaussian to the residual distribution. The vertex resolution as a function of the momentum is shown in Figure \ref{fig:resR} and Figure \ref{fig:resZ}

\begin{center}
\includegraphics[width=7.5cm]{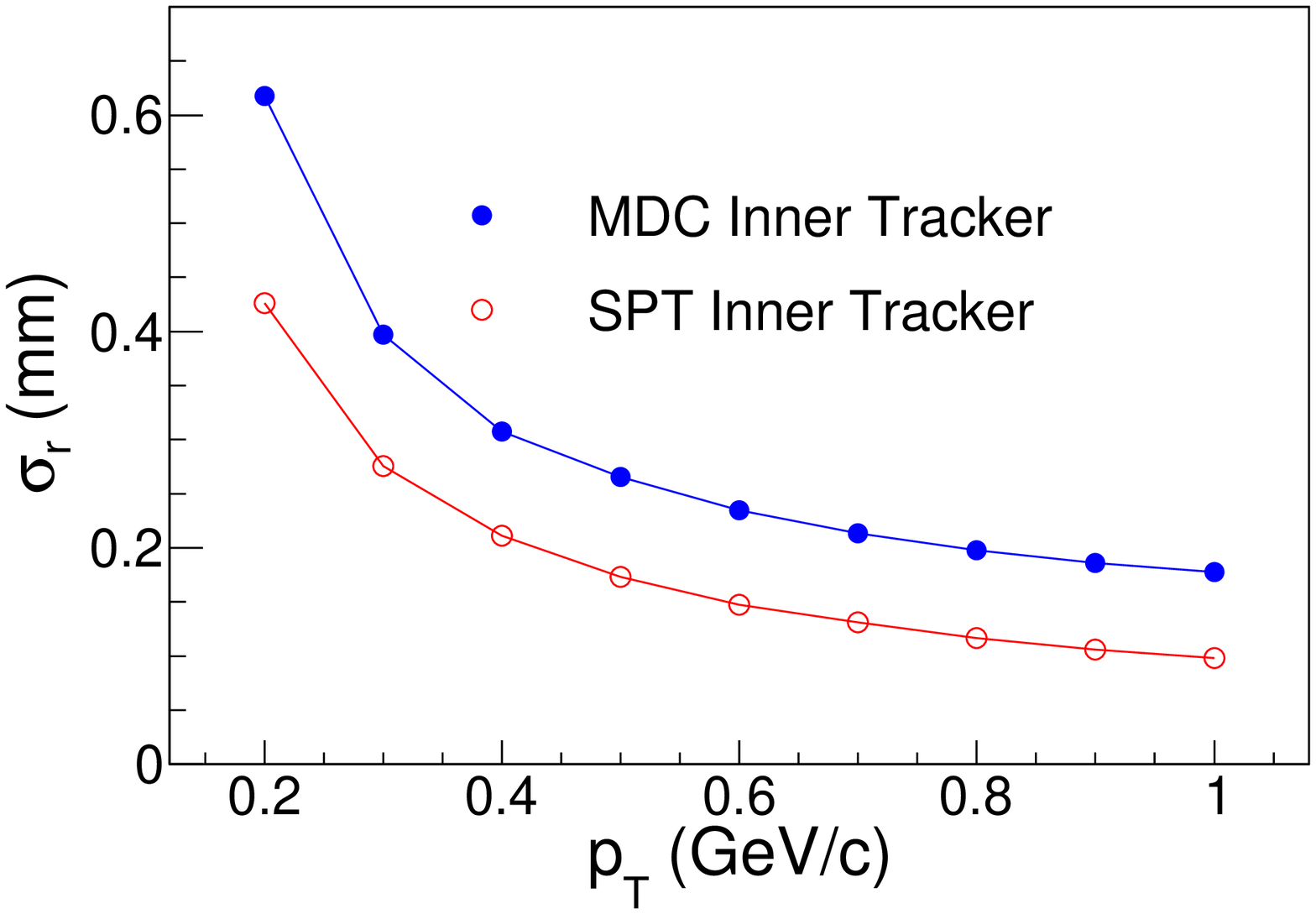}
\figcaption{\label{fig:resR}The vertex resolution in R as a function of the momentum}
\end{center}

\begin{center}
\includegraphics[width=7.5cm]{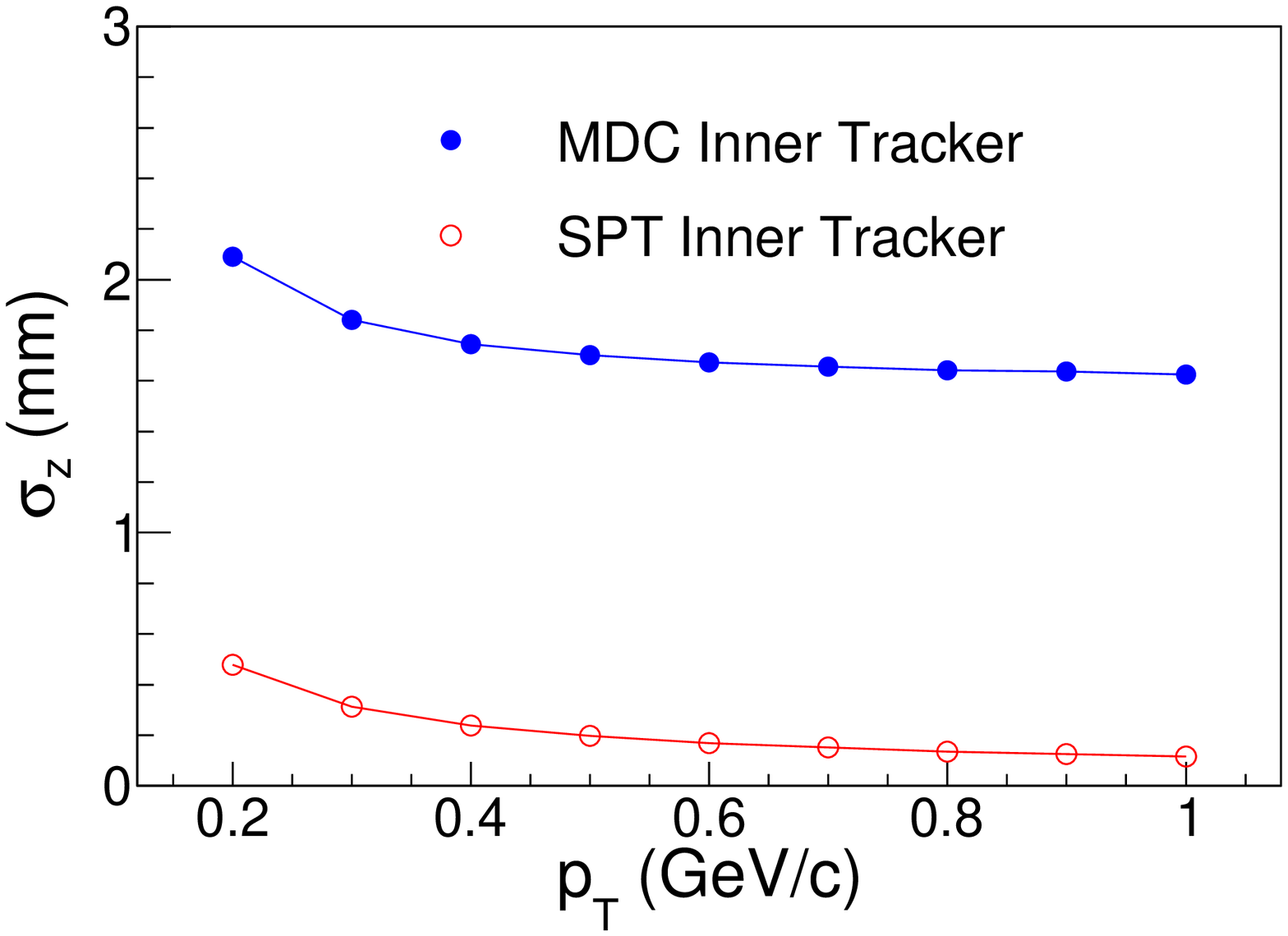}
\figcaption{\label{fig:resZ}The vertex resolution in Z as a function of the momentum}
\end{center}

It's clear that the vertex resolution of SPT is much better than that of the current MDC, especially in z direction, because of the high spatial resolution of the silicon pixel tracker. The spatial resolution of MDC is about 120$\mu m$ in $R-\Phi$ direction and about 3mm in z direction, however, the spatial resolution of SPT is about 10$\mu m$ both in $R-\Phi$ and z direction.

\subsection{The tracking efficiency}
The tracking efficiency is defined as
$$\varepsilon = N_{rec} / N_{MC}$$
where $N_{MC}$ is the number of charged Monte Carlo tracks. $N_{rec}$ is the number of good reconstructed tracks. There are two criteria to determine whether a track is good reconstructed. The first is that more than 80\% of the total found hits in the track are true hits. The second is more than 25\% of the total true hits of the track should be found.

\begin{center}
\includegraphics[width=7.5cm]{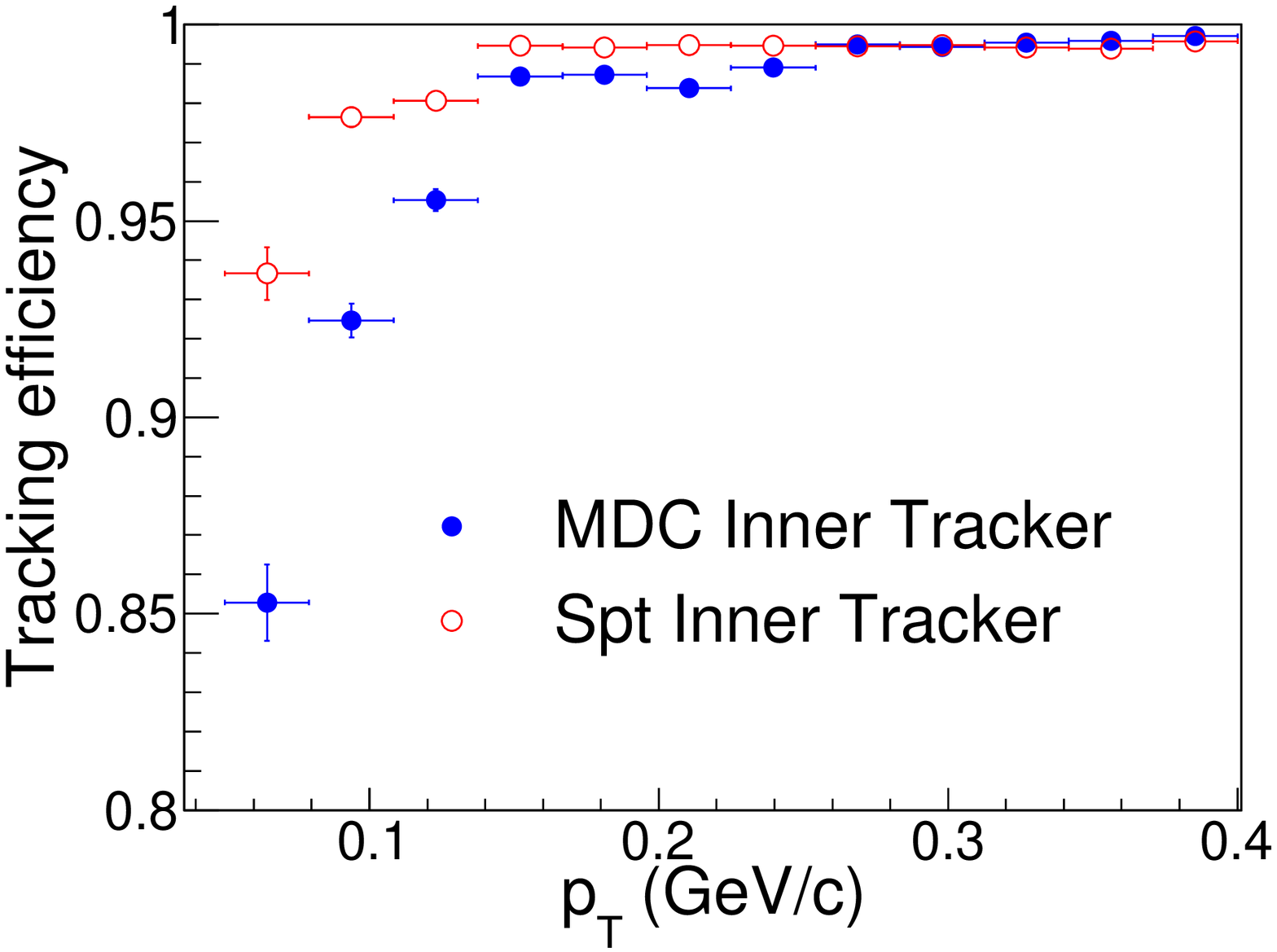}
\figcaption{\label{fig:trkEff}The tracking efficiency as a function of the transverse momentum}
\end{center}

The tracking efficiency as a function of the transverse momentum is shown in Figure \ref{fig:trkEff}. The tracking efficiency after applying the SPT inner tracker is similar with the current MDC in the high transverse momentum region. However, in the low transverse momentum region, the tracking efficiency will be significantly improved after applying the SPT inner tracker due to the high granularity and spatial resolution of the pixel detector.

\subsection{Check using $\Psi(2S) \to \pi^+ \pi^- J/\psi$}
The typical decay channel of $\Psi(2S) \to \pi^+ \pi^- J/\psi$, $J/\psi \to \mu^+ \mu^-$ is used to further verify the performance of SPT. The invariant mass distributions of $\mu^{+}\mu^{-}$ and the recoil mass distributions of $\pi^{+}\pi^{-}$ are shown in Figure \ref{fig:JpsiMass} and Figure \ref{fig:RecoilMass}.

\begin{center}
\includegraphics[width=7.5cm]{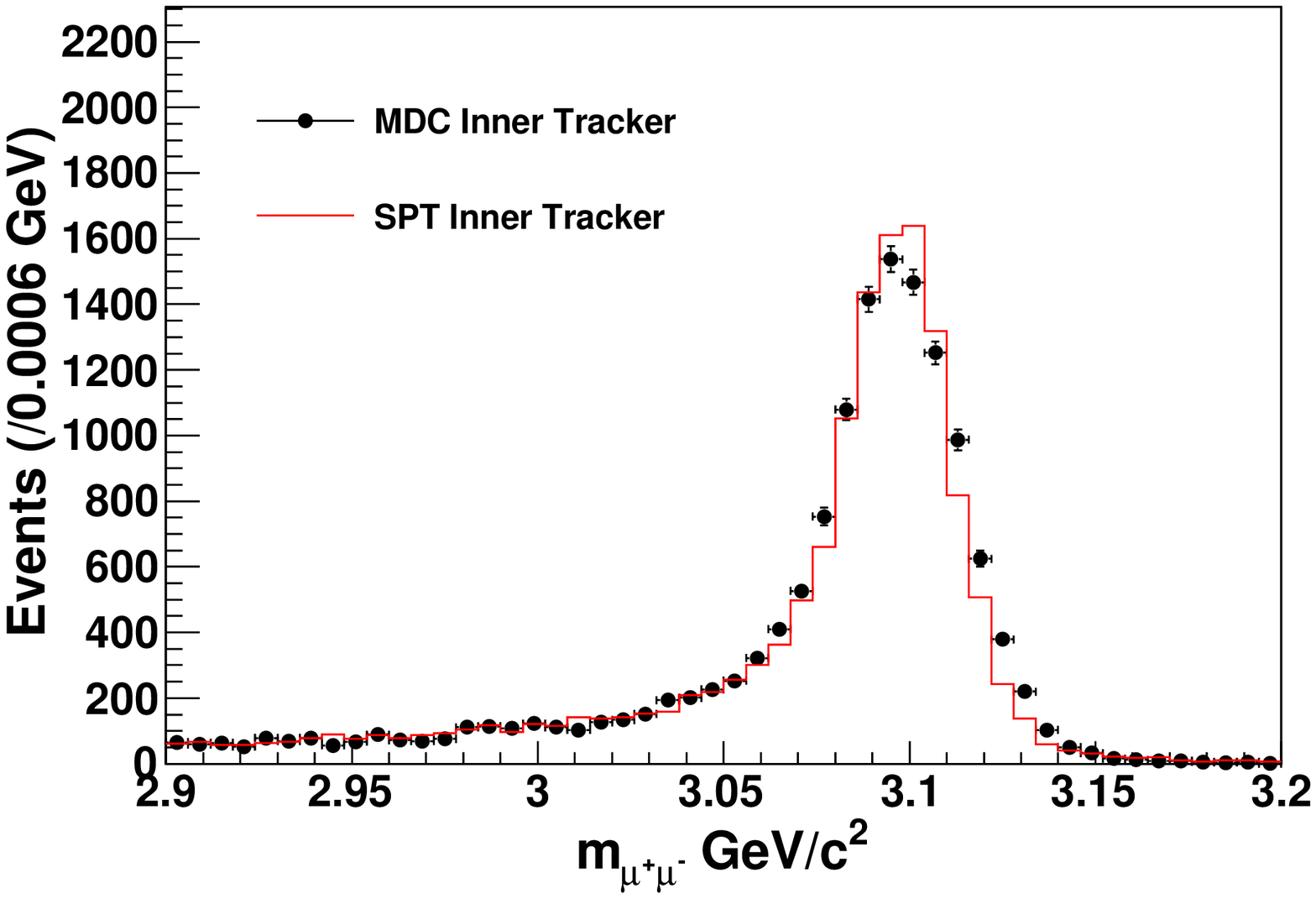}
\figcaption{\label{fig:JpsiMass} $\mu^+ \mu^-$ Invariant Mass}
\end{center}

\begin{center}
\includegraphics[width=7.5cm]{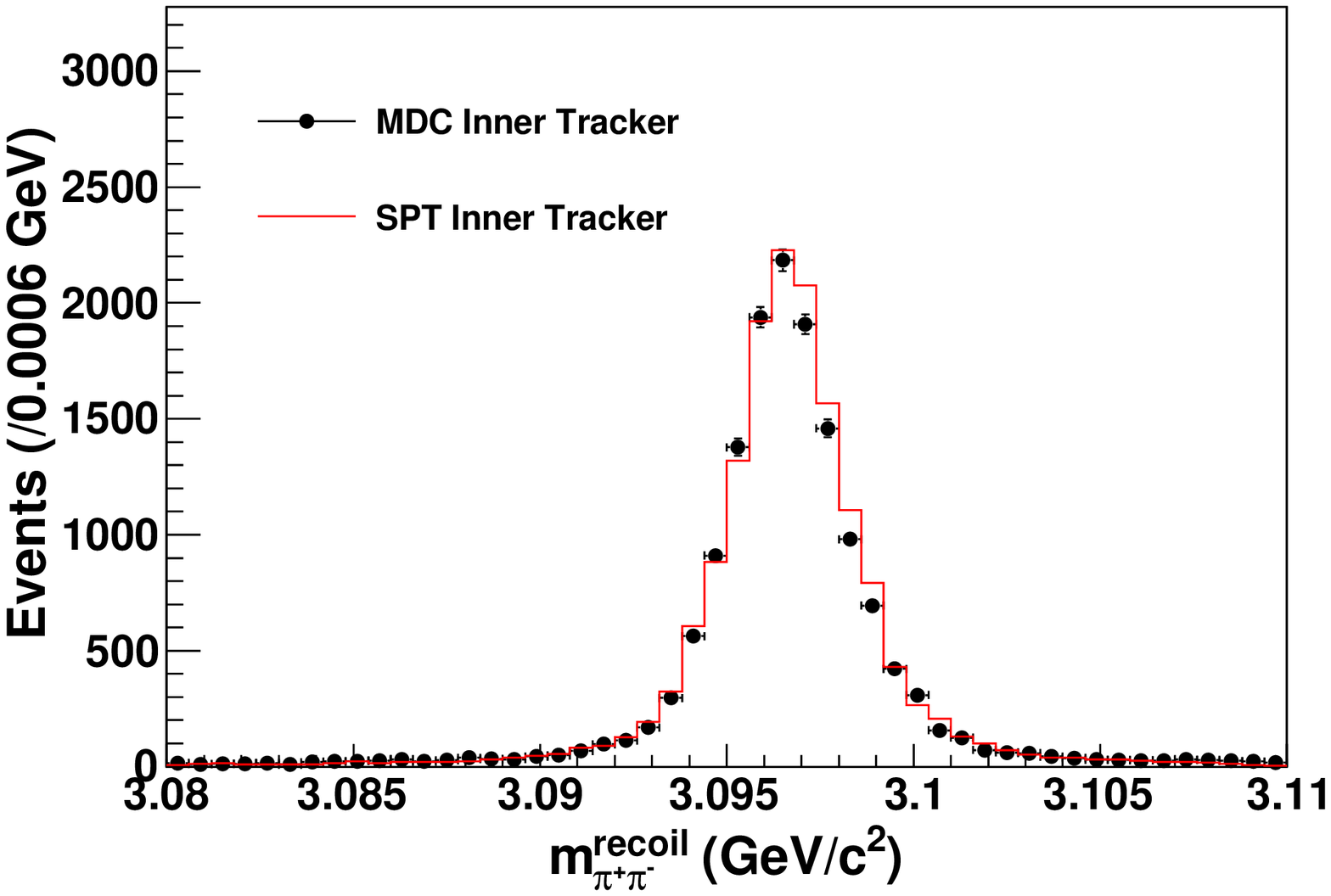}
\figcaption{\label{fig:RecoilMass} $\pi^+ \pi^-$ Recoil Mass}
\end{center}

The invariant mass resolution of $\mu^+ \mu^-$ obtained with SPT is a little better than that of MDC, which is mainly due to the improvement of momentum resolution at the high momentum region. However, the $\pi^+ \pi^-$ recoil mass distribution of SPT is not better than MDC because the momentum resolution for low momentum particle is not improved much due to multiple-scattering.

\section{Conclusion}
The inner chamber of MDC at BESIII is suffering from the aging problem due to the high beam related background. SPT is a very powerful candidate because of the high spatial resolution and radiation tolerance. In order to study the expected performance of SPT, the simulation of SPT based on Geant4 is implemented in BOSS and the track reconstruction software based on the combinatorial Kalman filter is developed. The results of the Monte Carlo study show that the momentum resolution, vertex resolution and the tracking efficiency are significantly improved if SPT inner tracker is applied. For instance, the momentum resolution for 1GeV/c tracks can be improved from 0.53\% to 0.46\%. The vertex resolution in R direction can be improved by 50\% and in Z direction can be improved from 0.16 cm to 0.013 cm at the momentum of 1 GeV/c.



\end{multicols}

\vspace{10mm}

\begin{multicols}{2}

\end{multicols}

\clearpage

\end{CJK*}
\end{document}